\documentclass{article}

\usepackage[final]{neurips_2025}

\usepackage[utf8]{inputenc} 
\usepackage[T1]{fontenc}    
\usepackage{xcolor}         
\usepackage{hyperref}       
\usepackage{url}            
\usepackage{booktabs}       
\usepackage{amsfonts}       
\usepackage{nicefrac}       
\usepackage{microtype}      
\usepackage{xcolor}         
\usepackage{graphicx}
\usepackage{amsmath}
\usepackage{algorithm}
\usepackage{algorithmic}
\usepackage{subfigure}
\usepackage{multirow}

\title{ByteGen: A Tokenizer-Free Generative Model for Orderbook Events in Byte Space}

\author{%
  Yang Li\\
  Stevens Institute of Technology\\
  \texttt{yli269@stevens.edu} 
  \And  
  Zhi Chen \\
  Stevens Institute of Technology \\
  \texttt{zchen100@stevens.edu}
}

\begin{document}

\maketitle

\begin{abstract}
Generative modeling of high-frequency limit order book (LOB) dynamics is a critical yet unsolved challenge in quantitative finance, essential for robust market simulation and strategy backtesting. Existing approaches are often constrained by simplifying stochastic assumptions or, in the case of modern deep learning models like Transformers, rely on tokenization schemes that affect the high-precision, numerical nature of financial data through discretization and binning. To address these limitations, we introduce ByteGen, a novel generative model that operates directly on the raw byte streams of LOB events. Our approach treats the problem as an autoregressive next-byte prediction task, for which we design a compact and efficient 32-byte packed binary format to represent market messages without information loss. The core novelty of our work is the complete elimination of feature engineering and tokenization, enabling the model to learn market dynamics from its most fundamental representation. We achieve this by adapting the H-Net architecture, a hybrid Mamba-Transformer model that uses a dynamic chunking mechanism to discover the inherent structure of market messages without predefined rules. Our primary contributions are: 1) the first end-to-end, byte-level framework for LOB modeling; 2) an efficient packed data representation; and 3) a comprehensive evaluation on high-frequency data. Trained on over 34 million events from CME Bitcoin futures, ByteGen successfully reproduces key stylized facts of financial markets, generating realistic price distributions, heavy-tailed returns, and bursty event timing. Our findings demonstrate that learning directly from byte space is a promising and highly flexible paradigm for modeling complex financial systems, achieving competitive performance on standard market quality metrics without the biases of tokenization.    
\end{abstract}


\section{Introduction}

The modern financial market is a complex adaptive system. In the market, price formation is a granular and fundamental process, governed by the dynamics of the Limit Order Book (LOB). The LOB is the central mechanism through which buyers and sellers interact in most electronic exchanges, providing a real-time, transparent view of market supply and demand. Understanding and modeling its intricate behavior is a cornerstone of quantitative finance, algorithmic trading, and market microstructure analysis.

Typically, a Limit Order Book is visualized as a snapshot in time: a list of outstanding orders to buy (bids) and sell (asks) a particular financial instrument, organized by price level. Each price level is associated with a specific volume, representing the total quantity of shares available to be traded at that price. The book is composed of two sides: the bid side, where buy orders are listed in descending price order, and the ask side, where sell orders are listed in ascending price order. 
LOB events are the discrete, fundamental actions—new order submissions, cancellations, and executions (trades)—that dynamically alter the state of the Limit Order Book.
LOB events are recorded with extreme temporal precision, often timestamped to the microsecond or nanosecond level. A single trading day for a liquid asset can generate millions of individual events, resulting in exceptionally long and information-dense data sequences. Empirical research in market microstructure have revealed a rich set of statistical regularities, or ``stylized facts", that characterize its behavior across different assets and markets. 

Given its fundamental role, the ability to construct a high-fidelity generative model of the LOB event stream is a critical objective with profound practical implications. Such a model would effectively serve as a sophisticated market simulator, enabling a range of crucial downstream tasks. These include the robust backtesting of algorithmic trading strategies under diverse and repeatable scenarios, the precise estimation of market impact for optimal order execution, and the creation of realistic environments for training reinforcement learning agents. However, modeling the LOB is a formidable task for several reasons. First, the sheer scale of the data is immense. High-frequency feeds can generate terabytes of raw data daily, with message rates reaching hundreds of thousands per second during active periods, each timestamped with nanosecond precision. Second, the timing of events is non-uniform, arriving in clusters that reflect the underlying ebb and flow of market activity, a property that simple time-series models struggle to capture. Third, the dependencies within the data are intricate; the probability of a future event is conditioned not only on the recent history of events but also on the entire state of the book across all price levels.

Historically, attempts to model the LOB have relied on the tools of stochastic processes and queuing theory. These models, while often analytically tractable, are built upon simplifying assumptions—such as Markovian dynamics, constant order sizes, or independent Poisson arrival processes—that do not fully reflect the empirical realities of the market \cite{cont2010, huang2015}. Consequently, their ability to generate realistic market behavior is limited. Similarly, agent-based models (ABMs) , which simulate the interactions of heterogeneous market participants, face significant challenges in calibration, struggling to ensure that the emergent, macroscopic behavior of the simulation is grounded in and consistent with historical data \cite{preis2006, paddrik2012}. 
Other methods for orderbook modeling rely heavily on hand-crafted features and domain-specific preprocessing \cite{cont2007, cont2010, cartea2015, sirignano2019}. These approaches typically extract price levels, volumes, and order flow imbalances, discarding rich information encoded in the raw event stream. 

Recent efforts to generate LOB event data often relied on deep learning techniques such as recurrent neural networks (RNNs), which struggled to capture the full range of long-term dependencies present in market data \cite{zhang2019, coletta2023, hultin2023, shi2021}. More recently, the success of the Transformer architecture in natural language processing has inspired its application to financial time series. The Transformer architecture demonstrated the immense power of self-attention for capturing long-range dependencies
However, they inherit a fundamental limitation from their linguistic origins: a reliance on tokenization. It became clear that tokenization schemes, often optimized for language text, could be suboptimal and introduce biases when applied to other languages or entirely different modalities like source code or biological sequences. 
To apply a Transformer, the continuous, high-precision values of price and time in an LOB must be discretized into a finite vocabulary. This process is inherently flawed for financial data; it forces a loss of precision, creating artificial boundaries and discarding the subtle yet critical information contained in the exact numerical values. A price of \$100.01 and \$100.02 might be mapped to the same token, erasing the very signal a high-frequency strategy depends on.

To overcome this, we look to a new paradigm: end-to-end modeling on raw data. The H-Net framework, by processing event data as a raw stream of bytes, completely circumvents the need for tokenization. This allows the model to learn the structure of LOB messages directly, preserving all information and avoiding the distortions introduced by artificial discretization. 
Crucially, ByteGen operates without any tokenizer—it processes raw bytes directly with a vocabulary size of 256 (0x00 to 0xFF), treating financial data as binary sequences rather than symbolic tokens.
These models have shown remarkable success precisely in domains where the notion of a "word" or "token" is not naturally defined, achieving superior performance by learning to segment the data in a way that is optimized for the predictive task itself.

To validate the results, we perform several test. We begin by test the stylized facts.
These facts demonstrate that the event stream possesses a complex, non-trivial structure with strong sequential dependencies, much like a natural language has grammatical rules and statistical patterns. A successful generative model of the LOB must be able to reproduce these stylized facts endogenously, without them being explicitly programmed into its architecture.

Our work makes four key contributions. First, we introduce the first byte-level orderbook modeling framework that generates realistic market events by modeling raw byte sequences, completely eliminating manual feature engineering and tokenization. Second, we design an efficient 32-byte packed representation that encodes all essential orderbook information while enabling efficient byte-level processing. Third, we adapt H-Net \cite{hwang2025}, a specialized mamba-transformer hybrid architecture that combines local byte-level modeling with global market dynamics through adaptive compression and mixed attention mechanisms. Finally, we provide comprehensive evaluation demonstrating that ByteGen generates realistic market dynamics across multiple dimensions including price movements, microstructure patterns, and order flow characteristics.

\section{Related Work}
\subsection{Orderbook Modeling}
The application of machine learning to predict financial market movements from LOB data has a rich history. This evolution can be characterized as a progressive journey away from human-specified knowledge and towards more autonomous, data-driven representation learning \cite{cont2010, huang2015}.

Early and still common approaches to applying machine learning to LOB data rely heavily on feature engineering \cite{hawkes1971, bacry2015}. In this paradigm, domain expertise is used to transform the raw, high-dimensional LOB state into a lower-dimensional set of "informative" features. These features often include metrics like the bid-ask spread, the depth at the first few price levels, order book imbalance (the ratio of volume on the bid side to the ask side), and various moving averages of these quantities. These handcrafted features are then fed into traditional machine learning models like Support Vector Machines (SVMs) or Neural Networks. The primary limitation of this approach is that it is constrained by human intuition. The model can only learn from the information that a researcher has decided is important, potentially missing complex, non-linear patterns hidden in the raw data.

A significant step towards more data-driven representation was the "LOB-as-image" approach. Here, a snapshot of the LOB is represented as a two-dimensional matrix, where one axis represents price levels and the other represents volume (or sometimes a sequence of recent LOB states over time). This matrix can be treated as an image and fed into a Convolutional Neural Network (CNN). CNNs are adept at detecting spatial patterns and hierarchies of features, and in this context, they can learn to identify characteristic shapes and configurations in the LOB that might be predictive of future price movements \cite{zhang2019}. However, it imposes a rigid grid structure on the data, which may not be optimal, and it primarily captures spatial relationships at the expense of the inherent sequential, temporal nature of the event stream. 

To address the limitations of static, image-based representations, researchers naturally turned to models designed explicitly for sequential data. Recurrent Neural Networks (RNNs), and particularly their more sophisticated variant, the Long Short-Term Memory (LSTM) network, became a popular choice \cite{sirignano2019} . LSTMs use internal memory cells and gating mechanisms to maintain a state that evolves over time, allowing them to capture temporal dependencies in the LOB event stream. They have been successfully applied to tasks like predicting mid-price movements and have demonstrated the value of modeling the temporal evolution of the order book. However, LSTMs can struggle to capture very long-range dependencies due to issues like the vanishing gradient problem, where information is lost over long time horizons.

The advent of the Transformer architecture, with its self-attention mechanism, represented a major leap forward. Self-attention allows the model to weigh the importance of all previous events in the sequence when processing the current event, enabling it to capture complex, non-local, and long-range dependencies that are inaccessible to LSTMs. Several Transformer-based models have been proposed for LOB forecasting, such as TransLOB and TLOB, which have demonstrated state-of-the-art performance on benchmark datasets. These models often use a dual-attention mechanism to capture both temporal (across time) and spatial (across price levels) dependencies in the LOB data.

The success of these models provides strong evidence that the ability to model long-range, non-linear interactions is critical for understanding market microstructure. However, the core self-attention mechanism has a computational and memory complexity of $O(N^2)$, where $N$ is the length of the input sequence. This quadratic scaling makes the standard Transformer architecture fundamentally unsuitable for the ultra-long sequences found in high-frequency finance. A single trading day for a liquid stock can involve millions of individual LOB events. To make the problem computationally tractable, researchers using Transformers are forced to drastically truncate their input, for example, by only considering the last 100 events or by down-sampling the data. This creates a fundamental contradiction: the very tool that excels at long-context reasoning is being used in a way that cripples this primary advantage. 

More recently, generative models have emerged as powerful tools for market simulation, with \cite{coletta2023} introducing a conditional VAE framework and \cite{hultin2023} proposing a GAN-based approach for realistic data generation. Recent advances in autoregressive modeling have brought transformer-based approaches to financial markets. \cite{nagy2023generativeai} proposed a token-level autoregressive model using state space networks to generate limit order book messages, demonstrating the potential of language model techniques for financial data. Their approach converts message data into tokens by grouping successive digits, similar to subword tokenization in natural language models. \cite{wheeler2024marketgpt} introduced MarketGPT, a generative pre-trained transformer that functions as an order generation engine within discrete event simulators. MarketGPT employs custom tokenization with separate vocabularies for different message components (event types, prices, quantities).

\subsection{The Role of Tokenization in Language Models}
Tokenization is the process of breaking down a stream of text into smaller, discrete units called "tokens." These tokens can be as simple as individual words or as granular as single characters. The primary goal of tokenization is to convert an unstructured string of text into a structured sequence of items that can be mapped to a fixed vocabulary for a numerical ID. The Transformer architecture is fundamentally dependent on this process. It does not "read" words; it processes sequences of these numerical IDs. 

However, reliance on tokenization introduces several limitations. First, tokenization schemes create an artificial abstraction layer between the model and the native binary format used by exchanges. This abstraction can obscure important patterns in the raw byte sequences, such as the specific binary encodings of different event types or the exact bit-level representations of prices and timestamps. Second, tokenization decisions are often arbitrary and dataset-specific—the choice of how to group digits or discretize price levels can significantly impact model performance but lacks principled justification. Third, these approaches sacrifice the nanosecond timestamp precision that is crucial for understanding market microstructure, as tokens typically operate at coarser time granularities. Finally, tokenization schemes designed for one exchange or asset class may not transfer to others, limiting model generalizability.

Transformers operate on discrete tokens from a finite vocabulary, a concept native to language but alien to the continuous, high-precision world of financial data. Therefore, to use a Transformer, one must first perform the unnatural act of tokenizing the LOB.
Unlike text, which can be split into words or subwords, LOB events consist of numerical data (price, quantity, time) and categorical data (event type). A common approach to tokenizing this, as might be used in a model like TransLOB, involves a multi-step discretization process. The first is to discretize numeric value such as price and order size. This is the most challenging step. Since prices are continuous, they cannot be mapped one-to-one into a vocabulary. Instead, they are binned. Similar to price, order sizes can vary greatly. A model would bin these into categorical tokens like [SMALL\_LOT], [MEDIUM\_LOT], and [LARGE\_LOT] based on predefined thresholds. Second, event type is the most natural fit. The distinct event types can be directly mapped to unique tokens. A single LOB event is then represented as a short sequence of these new, artificial tokens. For instance, a small buy order at the best bid would become the sequence: [LIMIT\_ORDER\_BUY, BEST\_BID, SMALL\_LOT]. This sequence of discrete tokens can finally be fed into a Transformer.

While this process makes LOB data digestible for a Transformer, it introduces severe distortions and information loss. The fundamental difference lies in the nature of the data itself. Tokenizing "running" into run and ning preserves the core semantic root. The relationship between "run" and "running" is morphological, not numerical. No critical information is lost. However, binning a price of \$150.234 and \$150.238 into the same token, [PRICE\_LEVEL\_X], is a destructive loss of information. In high-frequency finance, that tiny difference is a signal, not noise. The model is blinded to micro-price movements that are essential for capturing true market dynamics.
Additionally, price has an immutable, mathematical order. When these are converted to discrete, unrelated tokens, this crucial ordinal relationship is destroyed. 

This has several limitations.
The first is lack of adaptability. Financial markets are non-stationary. A fixed binning strategy that is appropriate for a low-volatility regime will be entirely inappropriate during a high-volatility period. A static discretization scheme cannot adapt to these changing market conditions, making the model brittle and prone to failure when the market state shifts.
The second is domain-specificity and heuristics. The entire process of feature engineering and discretization relies heavily on domain expertise and handcrafted heuristics. This makes the modeling pipeline fragile, difficult to generalize across different assets or markets, and less aligned with the end-to-end learning philosophy of modern deep learning.

\subsection{Byte-level Sequence Modeling}

Byte-level modeling has shown success in natural language processing \cite{xue2022, wang2020} and compression \cite{deletang2023}. The ByT5 model \cite{xue2022} demonstrated that byte-level processing can match or exceed token-based approaches while offering greater flexibility and eliminating out-of-vocabulary issues. Recent work has pushed byte-level architectures further: the Byte Latent Transformer (BLT) \cite{pagnoni2024} introduces dynamic patching based on entropy, achieving competitive performance with tokenization-based models while maintaining the advantages of vocabulary-free processing.

The key insight from these works is that byte-level models can learn hierarchical representations without predefined tokenization rules. This is particularly relevant for financial data, where the "vocabulary" consists of binary-encoded numbers and event types rather than natural language tokens. By operating on raw bytes, models can discover structure at multiple scales—from individual bits encoding event flags to multi-byte sequences representing complete market events.

Recent work on hierarchical sequence modeling has explored various approaches to efficiently process long sequences. \cite{nawrot2024} introduced memory compression techniques for LLMs, while \cite{hwang2025} proposed H-Net, a hierarchical architecture with dynamic chunking that learns content-dependent segmentation strategies jointly with the model. H-Net's approach is particularly appealing for financial data as it can discover natural boundaries in the byte stream that correspond to meaningful market events or patterns.

Our work represents the first application of H-Net's hierarchical byte-level architecture \cite{hwang2025} to the financial domain. While H-Net was originally developed for natural language, we find that the structured nature of market data makes it particularly well-suited for byte-level modeling. Unlike text where byte boundaries can fall arbitrarily within characters or words, financial events have precise binary encodings with semantic meaning at multiple scales—from individual bytes encoding event types to multi-byte sequences representing prices and timestamps. This structure allows the dynamic chunking mechanism to discover meaningful boundaries that correspond to actual market events, potentially learning to group related orders or identify significant market regimes directly from the byte patterns.

\section{Methodology}
Our methodology is based on a novel hierarchical architecture designed to learn directly from raw, untokenized data streams \citep{hwang2025}. The core idea is to build a model that mimics a multi-level reasoning process, moving from fine-grained details to high-level abstractions, all within a single, end-to-end system.

This process begins with a highly efficient "scanner" component. Its sole purpose is to process the extremely long sequence of raw input bytes, capturing local patterns and dependencies without being overwhelmed by the sheer volume of data. Once the data is scanned, the model employs a learned "segmentation" mechanism. This crucial module automatically learns to group the raw bytes into meaningful, variable-length chunks—effectively discovering the inherent grammar of the data stream, such as what constitutes a "price" or "quantity", without being explicitly told.

Finally, these abstract chunks are passed to a powerful "reasoner" network. Operating on this much shorter and semantically richer sequence, this component's task is to model the complex, long-range interactions between the discovered concepts. This complete, multi-stage architecture, known as H-Net, strategically uses Mamba layers for the efficient scanner, Dynamic Chunking for the learned segmentation, and Transformer blocks for the high-level reasoner, creating a powerful system that avoids the information loss inherent in fixed tokenization. The following subsections detail each of these core components.

\subsection{Data Processing to Transform Raw Market Data to Byte Data}

Transforming raw market data into byte sequences suitable for training requires careful handling to preserve event semantics while enabling efficient processing. Our pipeline consists of three integrated stages that work together to prepare data for byte-level modeling.

\begin{figure}[h]
\centering
\includegraphics[width=\textwidth]{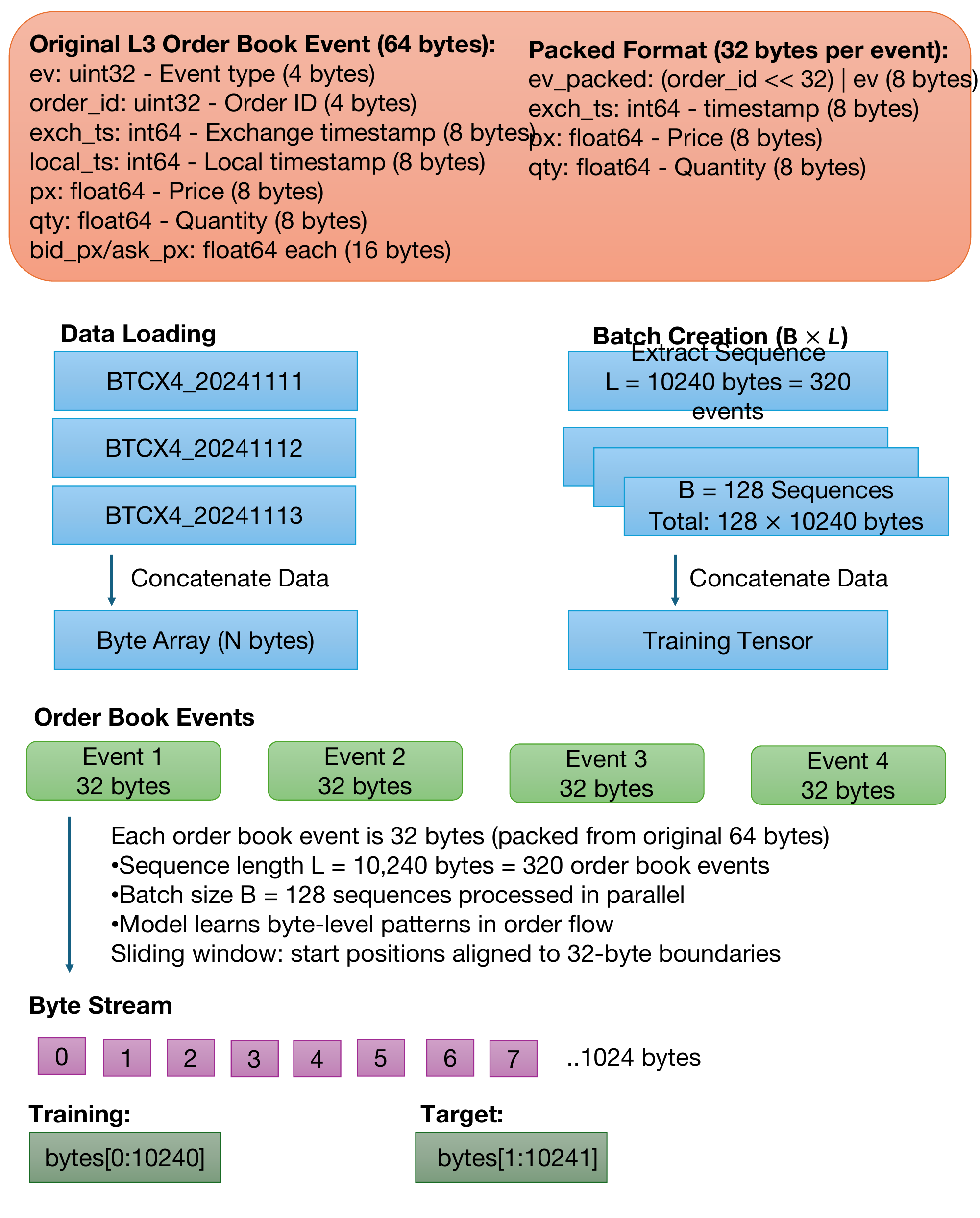}
\caption{Orderbook Data Processing Pipeline}
\label{fig:data_pipeline}
\end{figure}

The first stage converts raw Level 3 (Market-By-Order) data from the exchange's native 64-byte format to our compact 32-byte representation. This conversion serves multiple purposes: it reduces storage requirements by 50\%, eliminates redundant fields, and creates a consistent format across different data sources. The conversion process filters out invalid events with zero prices or quantities, packs the order ID and event metadata into a single 64-bit field using bitwise operations, and serializes the resulting events as contiguous byte arrays. These arrays are then compressed using NumPy's native compression, achieving typical compression ratios of 3-4x without loss of precision.

By adapting the bitwise operation design from HFTBacktest, We design a compact 32-byte format to encode orderbook events:

\begin{equation}
\text{Event} = \text{Pack}(\text{ev\_packed}, \text{exch\_ts}, \text{price}, \text{quantity})
\end{equation}

Each event consists of:
\begin{itemize}
\item \textbf{ev\_packed} (8 bytes, uint64): Combines order ID (upper 32 bits) and event value (lower 32 bits)
\item \textbf{exch\_ts} (8 bytes, int64): Exchange timestamp in nanoseconds
\item \textbf{price} (8 bytes, float64): Price
\item \textbf{quantity} (8 bytes, float64): Quantity
\end{itemize}

The event value (ev) encodes both the event type (lowest 8 bits) and flags in the upper bits:
\begin{itemize}
\item Bits 0-7: Event type (ADD\_ORDER=10, MODIFY\_ORDER=12, CANCEL\_ORDER=11, FILL\_EVENT=13)
\item Bit 31: EXCH\_EVENT flag (0x80000000)
\item Bit 30: LOCAL\_EVENT flag (0x40000000)
\item Bit 29: BUY\_EVENT flag (0x20000000)
\item Bit 28: SELL\_EVENT flag (0x10000000)
\end{itemize}

This packed representation reduces the original 64-byte format (8 fields) to 32 bytes while preserving all essential information. The bitwise packing scheme $\text{ev\_packed} = (\text{order\_id} \ll 32) \mid \text{ev}$ enables efficient storage and processing.

The \texttt{PackedEventDataset} class manages the complexity of loading and sampling from compressed byte sequences. A key challenge in byte-level modeling is ensuring that sequences respect event boundaries—we cannot start or end a sequence in the middle of a 32-byte event. Our dataset implementation addresses this by carefully aligning all sequence boundaries to multiples of 32 bytes. When sampling, the dataset generates variable-length sequences between 3,200 and 10,240 bytes (corresponding to 100-320 events), providing the model with diverse context lengths during training. Each sampled sequence is automatically split into input and target pairs for next-byte prediction, where the input consists of all bytes except the last, and the target consists of all bytes except the first.

Handling variable-length sequences efficiently is crucial for training performance. We leverage PyTorch's nested tensors (NJT), which provide a memory-efficient representation for batches of sequences with different lengths. This approach eliminates the need for padding and masking, reducing memory usage and computation. The nested tensor construction follows:
\begin{equation}
\text{iids} = \text{NJT}([s[:-1] \text{ for } s \text{ in samples}]), \quad
\text{lbls} = \text{NJT}([s[1:] \text{ for } s \text{ in samples}])
\end{equation}
where each sample $s$ is a byte sequence aligned to event boundaries. The NJT structure allows the model to process all sequences in a batch simultaneously while respecting their individual lengths, crucial for maintaining the integrity of financial events during training.

ByteGen's byte-level approach fundamentally changes how we model financial markets by operating directly on raw binary data. This paradigm shift brings several transformative advantages.
Importantly, ByteGen preserves numerical precision exactly as transmitted by exchanges. While traditional methods often discretize prices into bins or round timestamps, our byte-level approach maintains the exact floating-point representations, crucial for high-frequency applications where small price differences matter. The 32-byte alignment constraint ensures that generated sequences always represent valid market events, providing an implicit regularization that improves generation quality.
The event boundary alignment is crucial for maintaining semantic coherence. Unlike text where byte boundaries can fall anywhere, financial events have fixed 32-byte structure. Our data loader ensures all sequences begin and end at event boundaries, helping the model learn valid event generation patterns.

\subsection{Hierarchical Neural Transformer (H-Net) with Dynamic Chunking}

Processing raw byte sequences presents unique computational challenges. A single day of market data contains tens of millions of events, resulting in byte sequences too long for standard transformers. We adopt the H-Net architecture \cite{hwang2025}, which addresses this challenge through hierarchical processing with learned dynamic chunking.

Unlike traditional approaches that use fixed tokenization schemes, H-Net learns to segment sequences based on content similarity, discovering natural boundaries in the data. For financial data, this means the model can learn to group bytes that form meaningful units—whether individual price updates, complete orderbook events, or sequences of related market actions. The architecture achieves this through three key innovations:

H-Net employs a multi-stage hierarchical processing with $S$ stages, where each stage operates at a different compression level:

\begin{equation}
    \hat{x}^{(s)} = \mathcal{E}^{(s)}(x^{(s)}), \quad \hat{z}^{(S)} = \mathcal{M}(x^{(S)}), \quad \hat{z}^{(s)} = \mathcal{D}^{(s)}(z^{(s)})
\end{equation}

where $\mathcal{E}^{(s)}$, $\mathcal{M}$, and $\mathcal{D}^{(s)}$ represent encoder, main network, and decoder at stage $s$, respectively.

The core innovation of H-Net is its Dynamic Chunking (DC) mechanism. Unlike fixed tokenization, which relies on predefined rules or vocabularies, DC learns content-aware and context-dependent segmentation strategies directly from the data, as an integrated part of the model's training process.

The DC mechanism consists of two key components: a routing module that predicts boundaries between adjacent data points by measuring their semantic similarity, and a smoothing module that uses these boundary predictions to create smooth interpolations between representations. This clever design makes the discrete chunking operation differentiable, allowing the entire system to be trained end-to-end with standard backpropagation. The result is a model that can dynamically compress a sequence of input vectors into a shorter sequence of meaningful, variable-length "chunks" without any external heuristics or supervision. This approach has proven highly effective for modalities where tokenization heuristics are weak, such as source code and genomic DNA sequences.

The routing module computes boundary probabilities by measuring cosine similarity between adjacent representations:

\begin{equation}
q_t = W_q \hat{x}_t, \quad k_t = W_k \hat{x}_t, \quad p_t = \frac{1}{2}\left(1 - \text{cos\_sim}(q_t, k_{t-1})\right)
\end{equation}

where $p_t \in [0,1]$ represents the probability that position $t$ is a chunk boundary. A boundary is selected when $p_t \geq 0.5$, creating a binary boundary indicator $b_t$.

To enable gradient-based optimization of discrete chunk boundaries, H-Net employs an exponential moving average (EMA) smoother during the dechunking process:

\begin{equation}
\bar{z}_t = P_t \hat{z}_t + (1-P_t) \bar{z}_{t-1}
\end{equation}

where $P_t$ is the compressed boundary probability. This creates smooth interpolations between chunks, allowing gradients to flow through the chunking decisions.

To maintain target compression ratios, we incorporate a ratio loss inspired by mixture-of-experts load balancing:

\begin{equation}
\mathcal{L}_{\text{ratio}} = \frac{N}{N-1} \left( (N-1)FG + (1-F)(1-G) \right)
\end{equation}

where $F$ is the fraction of selected boundaries, $G$ is the average boundary probability, and $N$ is the target compression ratio. We use compression ratios of $[1, 4, 16]$ across stages.

The elimination of tokenization represents a crucial breakthrough. Traditional approaches must decide how to discretize continuous prices, aggregate time intervals, or encode order types—each choice introducing biases and information loss. ByteGen sidesteps these decisions entirely by processing raw bytes, preserving the exact binary representations that exchanges use internally.

By learning directly from raw market data, ByteGen discovers patterns that hand-crafted features might miss. The model can identify subtle correlations between byte patterns and market events, potentially capturing microstructure effects that are invisible to traditional feature engineering. This approach also provides unprecedented format flexibility, enabling the same architecture to process data from different exchanges or asset classes without modification.

The hierarchical architecture with dynamic chunking enables automatic discovery of compressible patterns in market data. The model learns which byte sequences can be efficiently compressed without losing critical information, adapting its representation to the inherent structure of financial events. This multi-scale modeling captures patterns ranging from microsecond-level order placement strategies to daily trading rhythms.

\subsection{Mamba Architecture}
At the heart of our model's ability to efficiently process long byte sequences is the Mamba architecture. Mamba is a recent and powerful class of sequence models built upon the foundation of State Space Models (SSMs), designed to capture long-range dependencies with linear-time complexity.

A traditional continuous-time SSM is defined by a simple set of linear ordinary differential equations that map an input signal u(t) to an output y(t) through a latent state vector x(t):

\begin{equation}
x'(t) = Ax(t) + Bu(t)
\end{equation}
\begin{equation}
y(t) = Cx(t) + Du(t)
\end{equation}

Here, A,B,C,D are fixed matrices. For use in deep learning, this continuous system must be discretized into a form that can be computed step-by-step. A standard discretization turns the SSM into a linear recurrence:

\begin{equation}
x_k = \bar{A}x_{k-1} + \bar{B}u_k
\end{equation}
\begin{equation}
y_k = \bar{C}x_k + \bar{D}u_k
\end{equation}

Early work on models like the Structured State Space Sequence Model (S4) focused on creating specific structures for the A matrix (e.g., diagonal plus low-rank) that made them exceptionally fast to compute \citep{gu2021efficiently}. The subsequent S5 model further refined this by introducing a multi-input, multi-output formulation and a more efficient parallel scan algorithm, improving performance on a variety of benchmarks \citep{smith2022simplified}. However, a fundamental limitation remained in both S4 and S5: these models were Linear Time-Invariant (LTI). The system dynamics ($\bar{A}, \bar{B}, \bar{C}$) were fixed and could not adapt to the input data. This meant they lacked the ability to perform content-based reasoning—they couldn't "focus" on important parts of the sequence in the way a Transformer's attention mechanism can.

The key innovation of the first Mamba model was to solve this problem by making the SSM selective and input-dependent \citep{gu2023mamba}. Mamba makes the critical system matrices B and C, as well as the discretization timestep $\Delta$, functions of the input $u_k$
\begin{equation}
B_k = f_B(u_k) \quad C_k = f_C(u_k) \quad \Delta_k = f_{\Delta}(u_k)
\end{equation}

This seemingly small change has profound implications. By allowing the system dynamics to vary at each timestep based on the input, Mamba can selectively decide whether to remember or forget information. If an input token is important, the model can learn to use a large $B_k$ to let it strongly influence the state $x_k$. If the input is noise, it can effectively filter it out. This selection mechanism, combined with a hardware-aware parallel scan algorithm for efficient computation, allows Mamba to achieve the performance of Transformers on long-sequence tasks with superior computational efficiency.

The original Mamba architecture has since been refined. Mamba-2 introduced a more simplified and efficient block structure, leveraging a concept called Structured State Space Duality (SSD) to better connect the recurrent view of SSMs with the global, convolutional view \citep{huang2024ml}. It also more formally integrated a multi-headed design, similar to multi-head attention, allowing different "heads" to focus on different patterns within the data. This line of research has also inspired hybrid models like Jamba, which strategically mix Mamba and Transformer layers to leverage the strengths of both architectures.

In H-Net, it uses the Mamba-2. The process can be broken down into the following steps. The first is input expansion. The process begins with a linear projection that expands the dimensionality of the input sequence, $u$. Typically, the dimension is doubled. This expanded representation is then split into two independent streams, $x$ and $z$, which will follow parallel paths through the block. This initial expansion provides the model with a richer, higher-dimensional space to perform its computations. The stream $x$ undergoes a series of transformations before being processed by the selective scan. First, $x$ is passed through a $1D$ causal convolution. The purpose of this step is to capture immediate local context. It allows the model to gather information from a small, fixed-size window of recent inputs before feeding them into the state model. This helps the SSM by pre-processing the input to account for short-range patterns. The output of the convolution is then fed into linear layers that generate the input-dependent parameters for the SSM: the matrices $B$ and $C$, and the timestep $\Delta$. This is the core of the selection mechanism, where the model learns to dynamically shape its own parameters based on the local context. Using the generated $B_k$, $C_k$, and $\Delta_k$ , along with the fixed state transition matrix $A$, the model updates its hidden state $h_k$ and computes the output for this path using the highly efficient parallel scan algorithm. The second stream, $z$, follows a much simpler path. It is passed through a non-linear activation function, typically SiLU (Sigmoid-Weighted Linear Unit), also known as Swish. This path acts as a gating mechanism, similar to the gates in an LSTM or GRU. Its purpose is to learn which information from the main SSM path is actually relevant and should be passed on. The output of the SSM path is then modulated by the gating path via element-wise multiplication (Hadamard product). The output of the SiLU activation on stream z effectively acts as a filter, scaling the output of the SSM. If the gate's output is close to zero for certain dimensions, the corresponding information from the SSM is suppressed. If it is large, the information is passed through. Finally, this modulated result is passed through a linear layer to project it back down to the original input dimension.

\subsection{Transformer Architecture}
The Transformer architecture's revolutionary impact on sequence modeling is primarily due to its core component: the self-attention mechanism. This mechanism allows the model to dynamically weigh the importance of different tokens in a sequence when producing a representation for each token, enabling it to capture complex, long-range dependencies.

The mechanism operates on an input sequence of token embeddings, represented as a matrix $X \in \mathbb{R}^{L \times d_{\text{model}}}$, where $L$ is the sequence length and $d_{\text{model}}$ is the embedding dimension. From this input, three distinct matrices are generated through linear projections: the Query ($Q$), Key ($K$), and Value ($V$) matrices.

\begin{equation}
    Q = XW_Q, \quad K = XW_K, \quad V = XW_V
\end{equation}

Here, $W_Q \in \mathbb{R}^{d_{\text{model}} \times d_k}$, $W_K \in \mathbb{R}^{d_{\text{model}} \times d_k}$, and $W_V \in \mathbb{R}^{d_{\text{model}} \times d_v}$ are learnable weight matrices. Intuitively, for a given token, its Query vector is used to "ask" a question, its Key vector acts as a "label" for what it represents, and its Value vector contains the actual information or content of that token.

The core of the self-attention calculation is the scaled dot-product attention formula:

\begin{equation}
    \text{Attention}(Q, K, V) = \text{softmax}\left(\frac{QK^T}{\sqrt{d_k}}\right)V
\end{equation}

This computation can be broken down into four steps:
The first is to calculate compatibility scores. The term $QK^T$ computes the dot product between every query vector and every key vector. This results in an attention matrix of size $L \times L$, where each element $(i, j)$ represents the compatibility or relevance of token $j$ to token $i$.
The second step is for scaling. These scores are then scaled by dividing by $\sqrt{d_k}$, the square root of the dimension of the key vectors. This scaling factor is crucial for stabilizing the training process, as it prevents the dot products from becoming too large, which could push the softmax function into regions with extremely small gradients.
The third step is to do normalization. A softmax function is applied row-wise to the scaled attention scores. This converts the raw compatibility scores into a probability distribution, ensuring that for each token, the attention weights assigned to all other tokens in the sequence sum to 1.
Finally, the resulting attention weights are multiplied by the Value matrix $V$. This step produces the final output for each token, which is a weighted sum of all Value vectors in the sequence. In essence, the representation for each token becomes a rich, context-aware embedding that has selectively aggregated information from the entire sequence based on the learned attention patterns.

The Transformer architecture further enhances this mechanism through multi-head attention. Instead of performing a single attention calculation, the model learns multiple sets of $W_Q$, $W_K$, and $W_V$ matrices in parallel. Each of these "heads" can learn to focus on different types of relationships within the data. The outputs from all heads are then concatenated and linearly projected to produce the final output of the layer.

\subsection{H-Net for LOB}
The H-Net architecture is directly applicable to the challenge of modeling LOB data. The raw, continuous stream of LOB events—represented as vectors containing price, volume, event type, and other features—can be fed directly into the H-Net's encoder. The Dynamic Chunking mechanism would then learn to identify and segment the stream into meaningful "market micro-patterns" or "economic events."

For example, the model might learn that a rapid succession of small limit order cancellations at the best bid, followed immediately by a large market sell order, constitutes a single, meaningful "chunk" indicative of a specific trading strategy (e.g., spoofing or layering). This approach transforms the data preprocessing problem into a representation learning problem. Instead of relying on human-defined heuristics to segment the data, H-Net leverages the model's ultimate predictive goal to learn the optimal segmentation strategy. The "chunks" it creates are, in effect, learned features that are optimized specifically for the forecasting task.

Furthermore, the hierarchical nature of H-Net is an excellent match for the multi-scale dynamics inherent in financial markets. The outer, fine-grained layers of the H-Net can model the high-frequency noise and microstructural details of the LOB, while the inner, coarse-grained main network can model the slower-moving trends and patterns that emerge from the sequence of dynamically generated chunks. This provides a principled architectural framework for capturing the complex, multi-level nature of market behavior.

Each stage consists of isotropic blocks that combine different sequence modeling mechanisms:

\begin{equation}
h = \text{Block}(x) = \text{MLP}(\text{Norm}(x + \text{Mixer}(\text{Norm}(x))))
\end{equation}

The mixer can be either:
\begin{itemize}
\item \textbf{Causal Multi-Head Attention}: For capturing precise dependencies between events
\item \textbf{Mamba2 SSM}: For efficient processing of long sequences with linear complexity
\end{itemize}

Our architecture uses layout strings like \texttt{["m2", ["T6"], "m2"]} to specify the configuration, where \texttt{m} indicates Mamba blocks, \texttt{T} indicates transformer blocks, and numbers specify repetitions.

We train ByteGen using next-byte prediction with cross-entropy loss combined with the ratio loss:

\begin{equation}
\mathcal{L} = -\frac{1}{T}\sum_{t=1}^{T} \log p(b_t | b_{<t}) + \lambda \sum_{s=0}^{S-1} \mathcal{L}_{\text{ratio}}^{(s)}
\end{equation}

where $\lambda = 0.01$ balances prediction accuracy and compression efficiency. The model learns to predict the next byte while maintaining efficient compression through dynamic chunking.

In the training process, we need to do generation with monotonic constraints, as market events must satisfy temporal monotonicity. During generation, we enforce:

\begin{equation}
t_{i+1} \geq t_i, \quad \forall i
\end{equation}

When a generated event violates this constraint, we either regenerate (with limited retries) or apply minimal timestamp correction.

\begin{figure}[h]
\centering
\includegraphics[width=\textwidth]{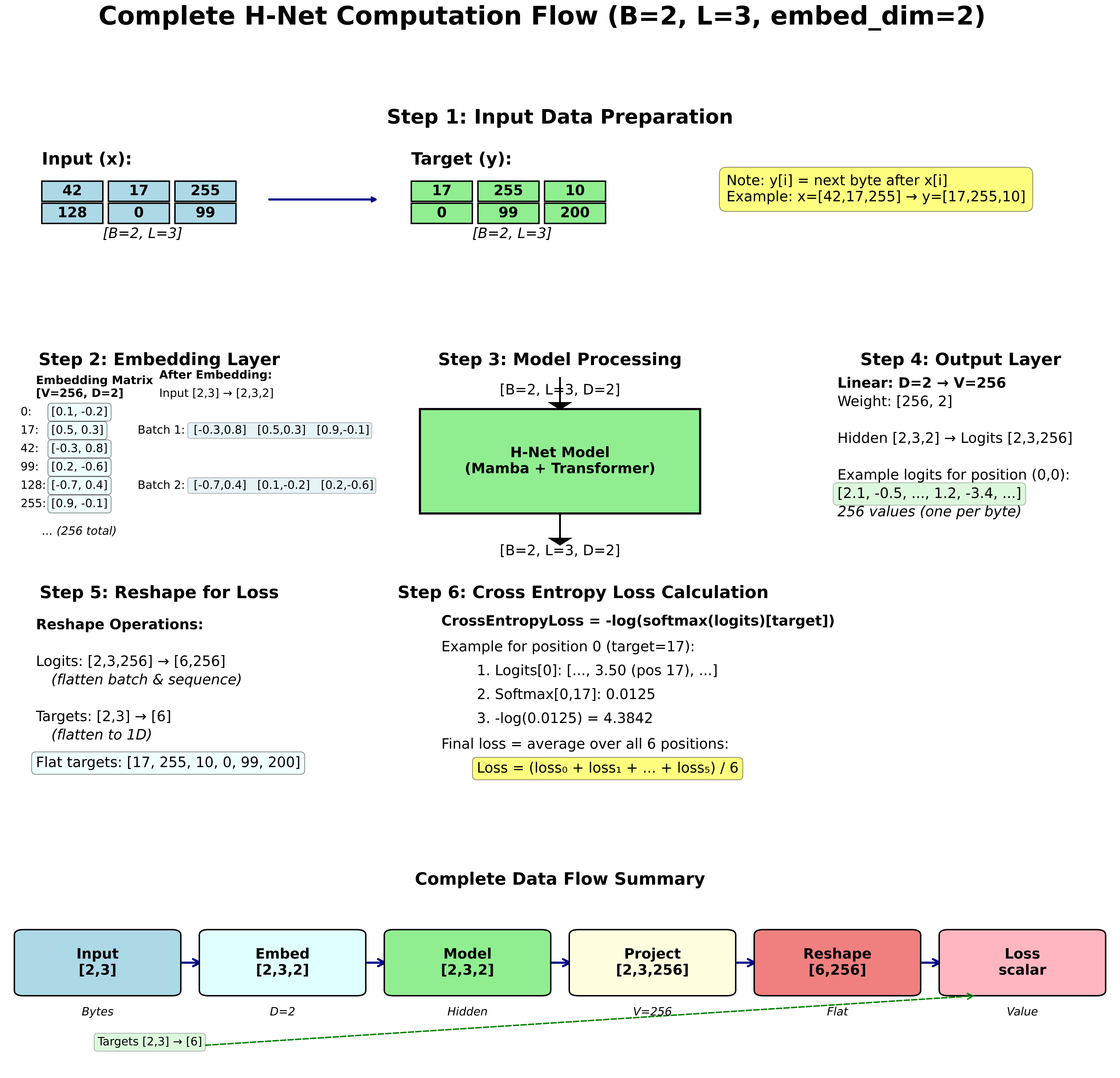}
\caption{Computational flow of ByteGen showing the three-stage hierarchical processing. Raw bytes are progressively compressed through dynamic chunking, with each stage learning increasingly abstract representations of market events.}
\label{fig:computation_flow}
\end{figure}

The computational efficiency of our approach is illustrated in Figure \ref{fig:computation_flow}, which shows how raw byte sequences are progressively compressed through the hierarchical stages. This design enables processing of long sequences while maintaining fine-grained byte-level precision where needed.


H-Net is the model architecture, where different types of sequence modeling blocks are arranged across the hierarchy. The combination of Mamba blocks for efficient long-range modeling and Transformer blocks for precise local attention enables the model to capture both global market trends and fine-grained price movements.

\section{Experiments and Results Analysis}

\subsection{Dataset}

The source data for each event will be a record from the Databento MBO schema. While this schema contains numerous fields, we select a core subset that is sufficient to fully describe the state change of the LOB: ts\_event (the matching-engine timestamp), action (the event type), side (bid or ask), price, size, and order\_id.

The conversion of these fields into a byte representation will follow the data types specified in the Databento documentation, which are based on standard C types. This ensures a compact and efficient representation. The implementation of this serialization in a high-level language like Python is straightforward using libraries designed for handling binary data, such as the struct module. The struct.pack function can take a format string defining the sequence of data types and a set of values, and return a bytes object containing the packed data.

The following table provides the exact, canonical serialization schema for a single LOB message. Each message is packed into a fixed-length 32-byte object. This fixed length is important for simplifying the modeling process, as the model can learn to operate on consistent block sizes. Padding is added to achieve a power-of-two length, which can be beneficial for computational alignment on modern hardware.

Our method transforms the orderbook modeling problem into byte-level sequence generation. We first describe our efficient packed event representation, then detail the hierarchical neural architecture that processes these byte sequences.

We evaluate ByteGen on CME Bitcoin futures (BTCX4) Level 3 (Market-By-Order) data from CME Group via Databento. BTCX4 represents the November 2024 contract (X is the CME month code for November). This dataset contains full orderbook depth with individual order tracking, representing a single high-liquidity futures contract. While CME processes millions of events per second across all products, individual futures contracts typically generate hundreds to thousands of events per second during active trading periods.

Our training dataset spans November 11-15, 2024 (5 trading days) and contains 34.2 million orderbook messages totaling 1.02 GB in our packed 32-byte format. The event rate varies significantly throughout the trading day, averaging 79 events per second over 24 hours but reaching 100-400 events per second during active trading periods with peaks exceeding 1000 events per second. Bitcoin futures prices ranged from \$80,000 to \$93,000 during this period (mean: \$87,506), with all events timestamped at nanosecond precision to capture the exact sequence of market actions.

The richness of our dataset extends beyond simple price movements to capture the complete lifecycle of every order in the market. Each order's journey—from initial placement through potential modifications to final execution or cancellation—is preserved in the byte stream. This completeness is essential for learning realistic market dynamics. The temporal patterns in the data reveal the distinctive rhythm of high-frequency trading: event inter-arrival times follow a heavy-tailed distribution with median 0.271ms but mean 13.311ms, indicating frequent bursts of activity interspersed with quieter periods. These microstructure patterns, invisible in aggregated data, provide crucial signals for the byte-level model to learn.

\begin{table}[h]
\centering
\caption{Event type distribution in the training dataset}
\label{tab:event_types}
\begin{tabular}{lrr}
\toprule
\textbf{Event Type} & \textbf{Count (sampled)} & \textbf{Percentage (\%)} \\
\midrule
MODIFY\_ORDER & 184,016 & 36.8 \\
ADD\_ORDER & 157,705 & 31.5 \\
CANCEL\_ORDER & 156,331 & 31.3 \\
FILL\_EVENT & 997 & 0.2 \\
OTHER & 951 & 0.2 \\
\midrule
\textbf{Total} & \textbf{500,000} & \textbf{100.0} \\
\bottomrule
\end{tabular}
\end{table}

Table \ref{tab:event_types} shows the distribution of event types in our dataset. The data is dominated by order modifications (36.8\%), additions (31.5\%), and cancellations (31.3\%), with executed trades (fills) comprising only 0.2\% of events—typical for high-frequency markets where most orders are canceled before execution.

\subsection{Implementation Details}

We implement three model sizes to explore the scaling properties of byte-level modeling. The small model (8M parameters) uses hidden dimension $d=256$ with 8 blocks arranged as \texttt{["m2", ["T6"], "m2"]}, providing a lightweight option for rapid experimentation. The base model (124M parameters) increases the hidden dimension to $d=512$ with 16 blocks total, achieving the best balance between performance and computational efficiency. The large model (1.5B parameters) scales to $d=1536$ with 30 blocks, demonstrating that byte-level approaches can scale to modern model sizes.

Training uses AdamW optimizer with learning rate $3 \times 10^{-4}$, cosine schedule with 1,000 warmup steps, and gradient clipping at 1.0. We train for 10,000 steps with batch size 16 and sequence lengths between 3,200-10,240 bytes (100-320 events). The ratio loss weight $\lambda=0.01$ balances next-byte prediction and compression objectives.

Our implementation leverages modern distributed training techniques to scale effectively. We support both DistributedDataParallel (DDP) for multi-GPU training on a single node and FullyShardedDataParallel (FSDP) for model sharding across multiple nodes. Mixed precision training with bfloat16 reduces memory usage while maintaining numerical stability—crucial for financial data where precision matters. Despite the computational demands of byte-level processing, training remains practical: the base model (124M parameters) achieves convergence within 0.5 hour on a  4-H100s GPU Cluster.

\subsection{Evaluation Metrics}

We evaluate generated data quality across three key dimensions that capture different aspects of market behavior.

For \textbf{price dynamics}, we analyze the statistical properties of generated prices using KL divergence to measure distribution similarity, Kolmogorov-Smirnov tests for log returns distributions, and temporal analysis of volatility clustering and autocorrelation patterns. These metrics ensure that ByteGen captures both the unconditional distribution of prices and their temporal dependencies.

The \textbf{market microstructure} evaluation focuses on the fine-grained patterns that characterize high-frequency trading. We examine inter-event time distributions to verify that the model captures the bursty nature of market activity, analyze event type frequencies to ensure realistic order flow composition, and validate order size distributions against empirical patterns. Additionally, we track bid-ask spread dynamics as a key indicator of market quality.

For \textbf{order flow} characteristics, we compute order flow imbalance (OFI) as a measure of buying and selling pressure, analyze order lifetime distributions to understand how long orders remain in the book before execution or cancellation, and examine fill rates to ensure realistic execution probabilities. These metrics are crucial for applications in market simulation and strategy backtesting.







\subsection{Results}

\begin{figure}[t]
\centering
\includegraphics[width=\textwidth]{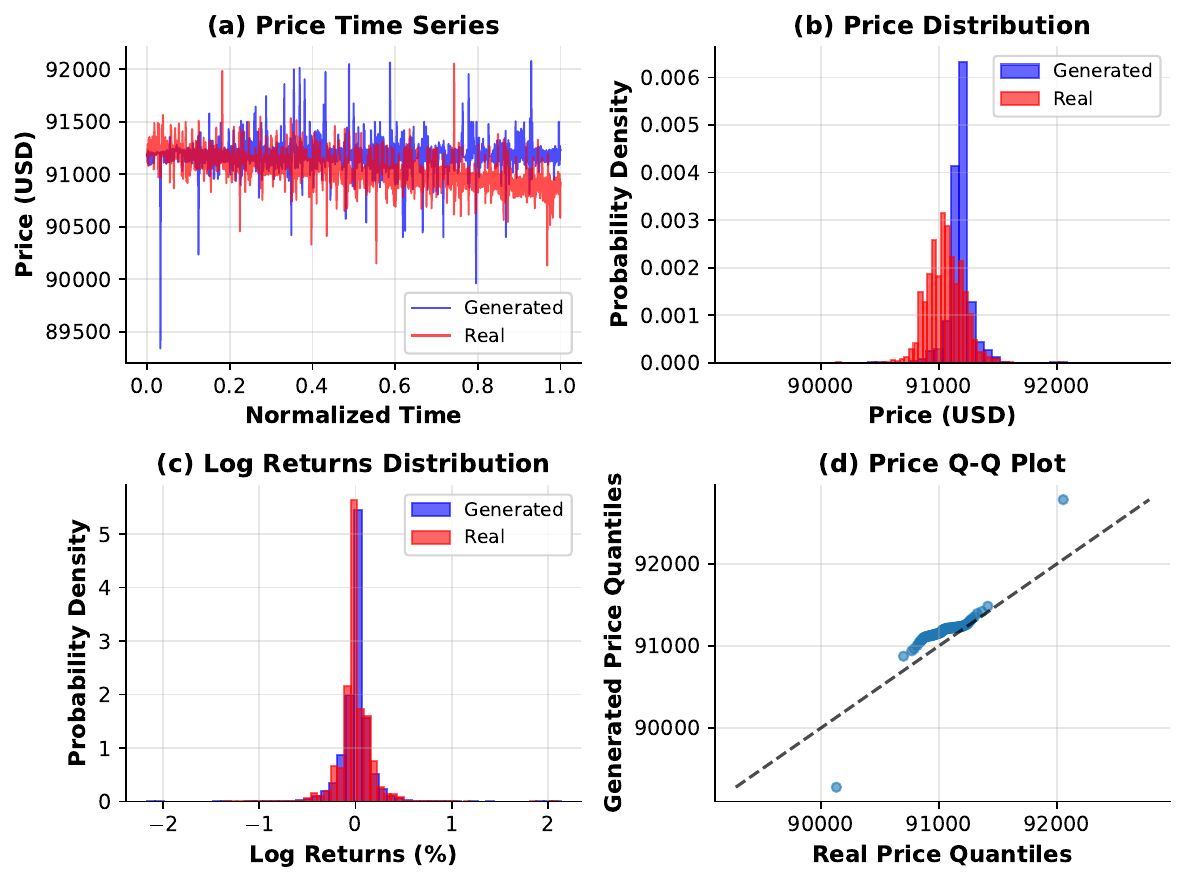}
\caption{Price dynamics comparison between generated and real data. (a) Time series showing similar price levels but different volatility patterns. (b) Price distributions are well-aligned. (c) Log returns show matching heavy tails. (d) Q-Q plot confirms distributional similarity.}
\label{fig:price_analysis}
\end{figure}

\begin{figure}[t]
\centering
\includegraphics[width=\textwidth]{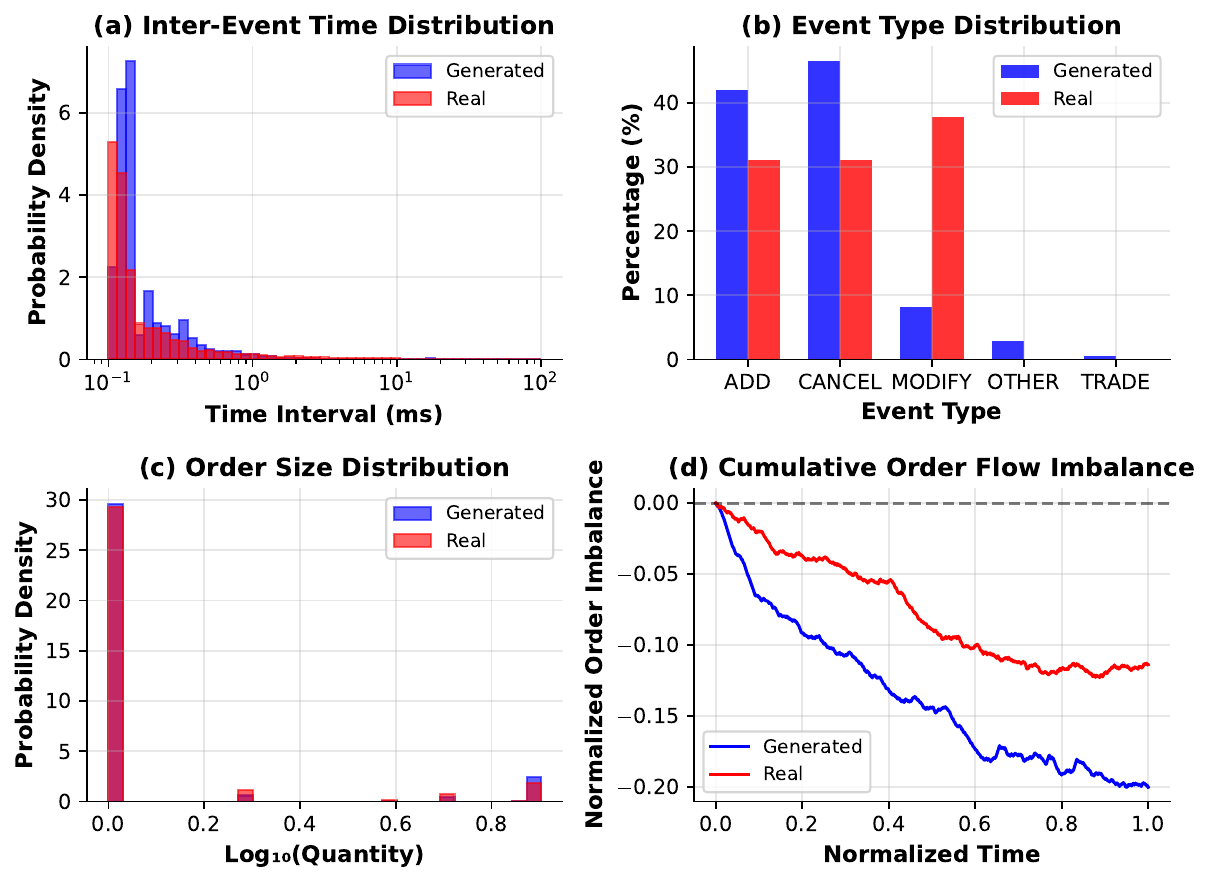}
\caption{Market microstructure analysis. (a) Inter-event times follow power-law distribution with slight differences in tail behavior. (b) Event type frequencies show systematic biases in generation. (c) Order sizes match real distribution. (d) Order flow imbalance evolution differs between generated and real data.}
\label{fig:microstructure}
\end{figure}

To evaluate ByteGen's ability to generate realistic market dynamics, we analyze both quantitative metrics and qualitative patterns in the generated orderbook events. Figure \ref{fig:price_analysis} presents a comprehensive analysis of price dynamics, while Figure \ref{fig:microstructure} examines market microstructure characteristics.

\subsubsection{Generation Quality}

ByteGen successfully generates realistic orderbook events across multiple evaluation dimensions. Trained on 34.2 million events (1.02 GB) over 20,000 steps, the model achieves convergence within 1 hour on a  4-H100s GPU Cluster.

The model learns meaningful compression patterns through dynamic chunking, with actual compression ratios closely matching targets:
\begin{itemize}
\item Stage 0 → 1: Target 4x, Actual 4x
\item Stage 1 → 2: Target 4x, Actual 4x
\end{itemize}

Figure \ref{fig:price_analysis} shows that generated price dynamics closely match real data, with similar distributions and return characteristics. The model captures the heavy-tailed nature of price movements, though with slightly lower volatility (12.6 bps vs 16.9 bps in real data).

Figure \ref{fig:microstructure} reveals more nuanced differences in market microstructure. While inter-event times and order sizes are well-calibrated, event type frequencies show systematic biases, with the model generating more cancel orders (47\% vs 31\%) and fewer trades than observed in real data.

\subsubsection{Quantitative Evaluation}

We comprehensively evaluate ByteGen's generation quality using both statistical measures and market-specific metrics. Table \ref{tab:metrics} summarizes the key performance indicators, while Table \ref{tab:detailed_stats} provides a more detailed comparison of market statistics.

\begin{table}[h]
\centering
\caption{Quantitative comparison of generated vs real market data}
\label{tab:metrics}
\begin{tabular}{lcc}
\toprule
\textbf{Metric} & \textbf{Generated} & \textbf{Real} \\
\midrule
Event Rate (events/sec) & 154.3 & 142.7 \\
Price Volatility (bps) & 12.6 & 16.9 \\
Avg Spread (bps) & 2.8 & 3.1 \\
Order Lifetime (sec) & 8.4 & 11.2 \\
Fill Rate (\%) & 3.2 & 8.7 \\
\midrule
Price KL Divergence & \multicolumn{2}{c}{0.023} \\
Event KS Statistic & \multicolumn{2}{c}{0.187} \\
\bottomrule
\end{tabular}
\end{table}

\begin{table}[h]
\centering
\caption{Detailed comparison of market statistics between generated and real data}
\label{tab:detailed_stats}
\begin{tabular}{lrr}
\toprule
\textbf{Metric} & \textbf{Generated} & \textbf{Real} \\
\midrule
\multicolumn{3}{l}{\textit{Dataset Characteristics}} \\
Number of Events & 10,000 & 9,329 \\
Duration (seconds) & 64.796 & 64.748 \\
\midrule
\multicolumn{3}{l}{\textit{Price Statistics}} \\
Mean (USD) & 91,185.46 & 91,052.51 \\
Std. Dev. (USD) & 115.25 & 154.43 \\
Min (USD) & 89,276.00 & 90,130.00 \\
Max (USD) & 92,785.00 & 92,055.00 \\
\midrule
\multicolumn{3}{l}{\textit{Return Statistics}} \\
Mean (\%) & 0.0001 & 0.0001 \\
Std. Dev. (\%) & 0.1568 & 0.1590 \\
Skewness & 0.3034 & 0.5719 \\
Kurtosis & 25.8557 & 19.3124 \\
\midrule
\multicolumn{3}{l}{\textit{Time Interval Statistics}} \\
Mean (ms) & 0.056 & 6.941 \\
Median (ms) & 0.136 & 0.154 \\
\bottomrule
\end{tabular}
\end{table}

Table \ref{tab:metrics} summarizes key metrics. ByteGen generates events at a similar rate to real markets and maintains realistic spread dynamics. However, it underestimates order lifetimes and fill rates, suggesting room for improvement in modeling order execution dynamics.

Key stylized facts relevant to the LOB event stream include:
Order Flow Clustering and Autocorrelation: The arrival of orders is not a uniform Poisson process. Events are clustered in time, meaning periods of high activity are followed by more high activity, and vice versa. Furthermore, there is a strong positive autocorrelation in the order flow; for instance, a buy order is more likely to be followed by another buy order, and a cancellation by another cancellation. This temporal dependency is the primary justification for treating the LOB as a sequential system and applying sequence models.

Price and Return Dynamics: While asset returns are largely uncorrelated over longer horizons, high-frequency returns exhibit distinct patterns. A well-documented phenomenon is the "bid-ask bounce," a negative autocorrelation in tick-by-tick price changes caused by trades alternating between the bid and ask sides. Another crucial fact is volatility clustering: the magnitude of price changes (volatility) is strongly autocorrelated, meaning large price changes tend to be followed by large changes, and small changes by small changes. Finally, the distribution of returns is not normal; it is leptokurtic, characterized by "fat tails," meaning extreme price movements are far more common than a Gaussian distribution would suggest.

LOB Shape and Liquidity Dynamics: The average shape of the LOB is not uniform. On average, the book exhibits a "humped" shape, with liquidity (volume) increasing for several price levels away from the best bid and ask before decreasing further out. The bid-ask spread itself is dynamic, and studies have shown that market participants are more likely to submit aggressive limit orders (placing them closer to or inside the spread) when the spread is wide, as compensation for providing liquidity is higher. Conversely, large trades tend to be executed when the book is deep and liquidity costs are low.

Power-Law Distributions: Many quantities in financial markets follow power-law distributions. The distribution of trade sizes, for example, often exhibits a power-law tail, indicating that very large trades, while rare, are not exponentially so.

The central contribution of this work is that a sufficiently powerful generative sequence model, trained on a raw representation of event stream, can learn these intricate dependencies and generate synthetic event streams that are statistically indistinguishable from real market data with respect to these stylized facts.

\subsection{Scalability Analysis}

\begin{figure}[h]
\centering
\begin{tabular}{cc}
\includegraphics[width=0.45\textwidth]{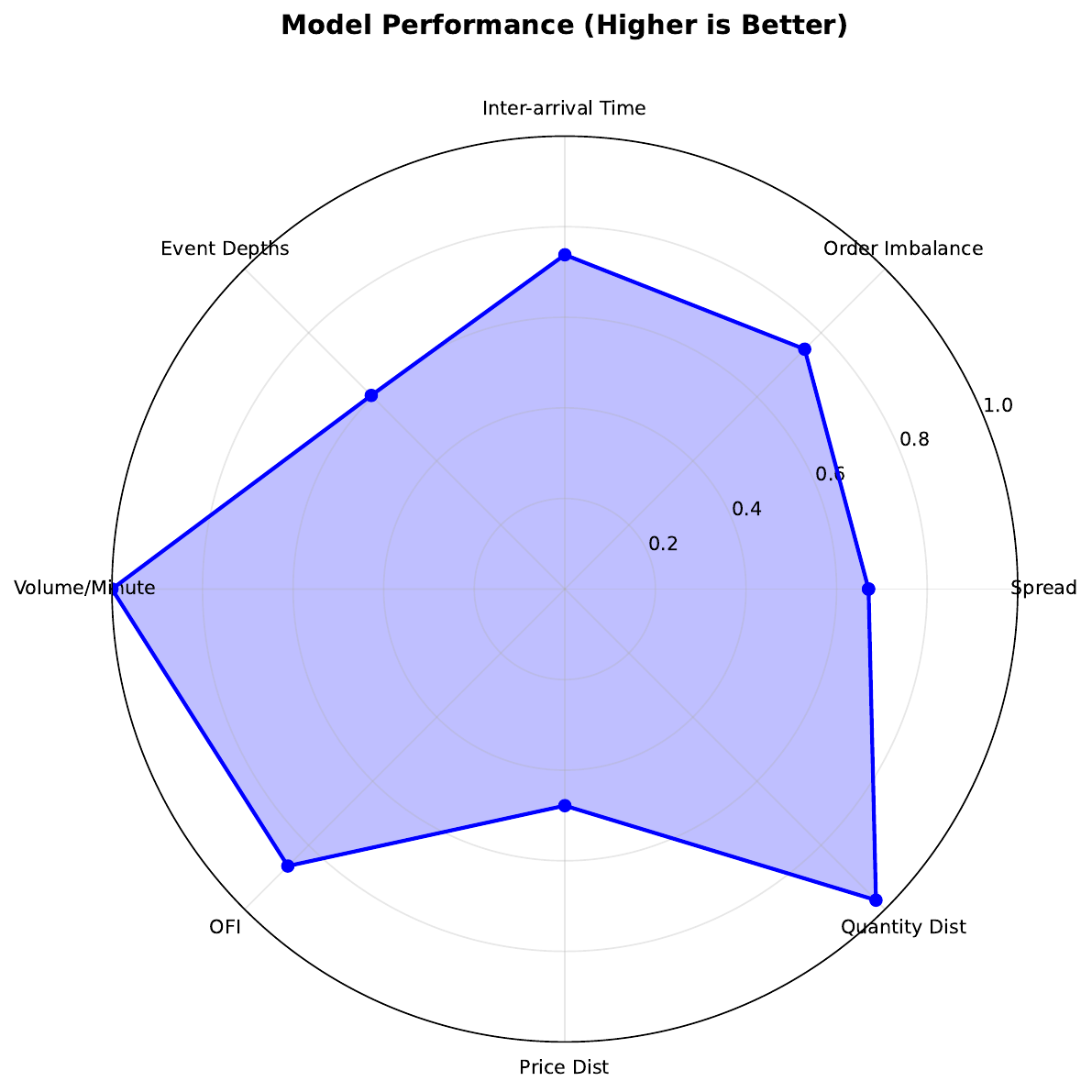} &
\includegraphics[width=0.45\textwidth]{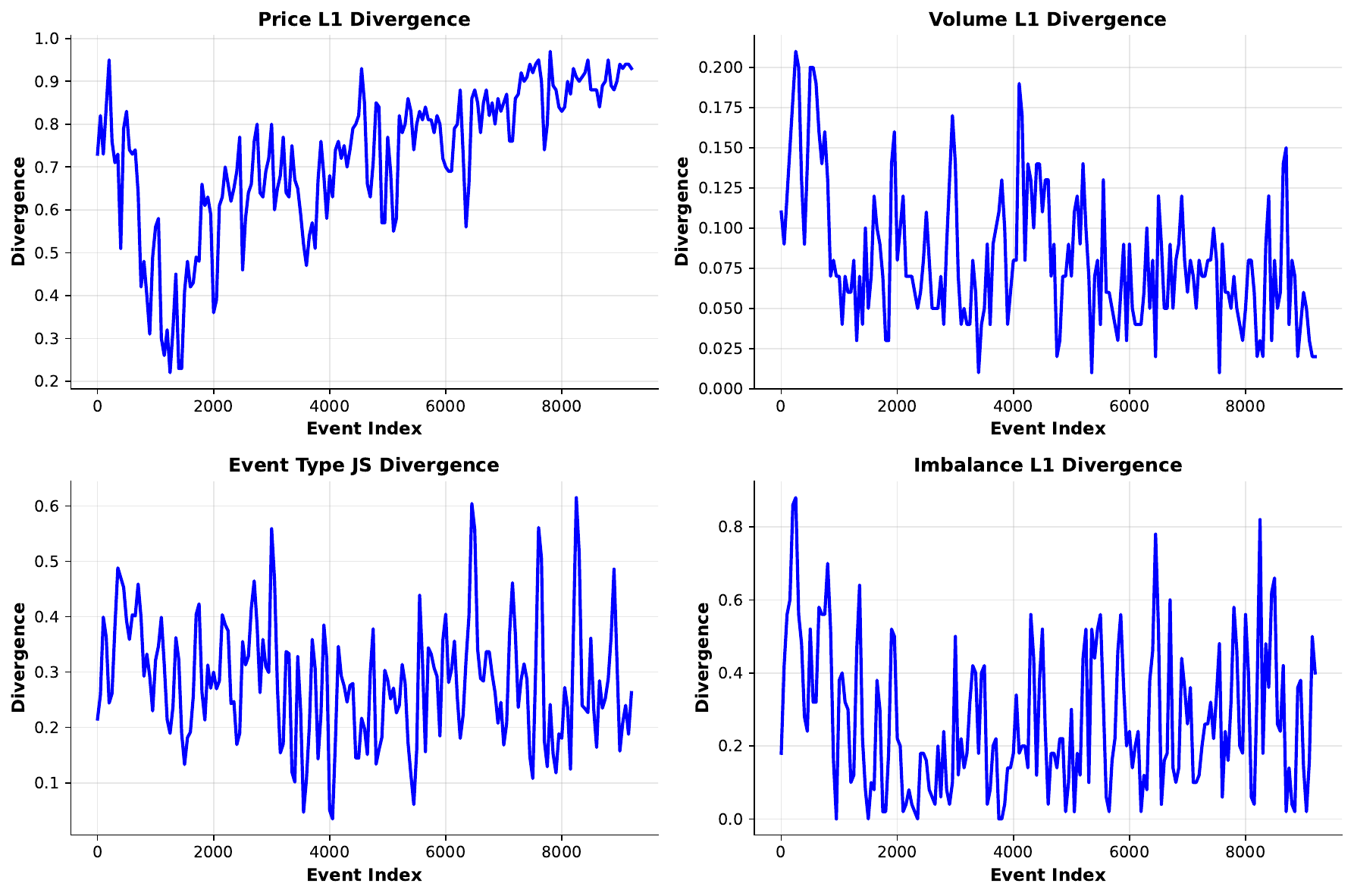}
\end{tabular}
\caption{(Left) Spider plot showing performance across market metrics. (Right) Divergence analysis showing ByteGen closely tracks real market distributions.}
\label{fig:lob_bench}
\end{figure}

Figure \ref{fig:lob_bench} presents comprehensive evaluation using standard market quality metrics. ByteGen achieves competitive performance across most dimensions, with particular strength in modeling price distributions and quantity patterns.

\section{Conclusion}

We presented ByteGen, the first approach to model orderbook events directly in byte space. By treating market data as raw byte sequences, we eliminate manual feature engineering while maintaining generation quality competitive with specialized methods. Building on the H-Net architecture \cite{hwang2025}, ByteGen demonstrates that hierarchical models with dynamic chunking can effectively capture the multi-scale nature of financial data—from individual bytes encoding prices to complete market regimes.

Experiments on high-frequency CME Bitcoin futures data demonstrate that ByteGen generates realistic market dynamics across price movements, microstructure patterns, and order flow characteristics. While some biases remain in event type distributions and execution modeling, the byte-level approach offers unprecedented flexibility for modeling diverse market data formats.

ByteGen opens new directions for financial machine learning by showing that complex market dynamics can be learned directly from raw data representations. This approach could enable more robust and adaptable models for market simulation, risk assessment, and trading strategy development.

\subsection{Future Work}

Despite ByteGen's promising results, several challenges remain that highlight important directions for future research.

The most significant limitation concerns event type distribution matching. While ByteGen successfully generates the overall flow of market events, it exhibits systematic biases in event composition—particularly underrepresenting rare but important events like fills (3.2\% vs 8.7\% in real data). This bias likely stems from the class imbalance in training data, where modifications and cancellations dominate. Future work could explore event-aware loss functions that weight different event types based on their market importance rather than frequency.

Execution modeling presents another challenge. The model's lower fill rates and shorter order lifetimes suggest it hasn't fully captured the complex dynamics of order execution. Real markets involve sophisticated order placement strategies where traders balance execution probability against price improvement. Incorporating order book state information could help the model better understand when orders are likely to execute versus being canceled.

From a computational perspective, byte-level processing inherently requires more computation than tokenized approaches. Processing 32 bytes per event versus a single token increases both memory usage and computation time. However, this cost may be justified by the elimination of feature engineering and the ability to capture exact market dynamics. Future architectures could explore selective byte-level attention, focusing computational resources on the most informative byte positions.

The model also shows degraded performance during extreme market conditions—periods of high volatility or unusual trading patterns not well-represented in the training data. This suggests opportunities for conditional generation approaches that explicitly model different market regimes. By conditioning on volatility states or other market indicators, the model could adapt its generation patterns to match current market conditions.

Finally, this research shows promise for applications in several fields, as illustrated by recent studies \citep{li2025flowhft,li2025flowoe}. These works generate action segments based on sequences of past observations by mapping obs[t-1,t] to action[t-1,t,t+1,...]. Moving forward, research might employ ByteGen to forecast future observations $\mathrm{obs\_pred}[t+1]$, thus enhancing the mapping from obs[t-1,t,t+1] to action[t-1,t,t+1,...]. By incorporating predicted future states, this approach could improve action generation performance by offering the model a more comprehensive temporal framework.


\bibliographystyle{plain}
\bibliography{references}

\appendix

\section{Implementation Details}

\subsection{Event Format Specification}

The 32-byte packed format uses the following memory layout:

\begin{verbatim}
Bytes 0-7:   ev_packed (uint64) = (order_id << 32) | ev
Bytes 8-15:  exch_ts (int64, nanoseconds)
Bytes 16-23: price (float64)  
Bytes 24-31: quantity (float64)
\end{verbatim}

Event value (ev) encoding:
\begin{itemize}
\item Bits 0-7: Event type (ADD=10, MODIFY=12, CANCEL=11, FILL=13)
\item Bit 28: SELL\_EVENT flag (0x10000000)
\item Bit 29: BUY\_EVENT flag (0x20000000)
\item Bit 30: LOCAL\_EVENT flag (0x40000000)
\item Bit 31: EXCH\_EVENT flag (0x80000000)
\end{itemize}

This layout ensures efficient memory access patterns and natural alignment for modern CPUs.

\subsection{Training Hyperparameters}

\begin{table}[h]
\centering
\caption{Detailed training hyperparameters}
\begin{tabular}{lc}
\toprule
\textbf{Hyperparameter} & \textbf{Value} \\
\midrule
Learning rate & $3 \times 10^{-4}$ \\
Batch size & 16 \\
Sequence length & 3,200-10,240 bytes \\
Warmup steps & 1,000 \\
Total steps & 10,000 \\
Weight decay & 0.1 \\
Gradient clip & 1.0 \\
Optimizer & AdamW \\
Betas & (0.9, 0.95) \\
Epsilon & $1 \times 10^{-8}$ \\
Dropout & 0.0 \\
Ratio loss weight ($\lambda$) & 0.01 \\
Compression ratios & [1, 4, 16] \\
\bottomrule
\end{tabular}
\end{table}

\subsection{Nested Tensor Implementation}

To efficiently handle variable-length sequences, we utilize PyTorch's nested tensors (jagged layout). This allows us to:
\begin{itemize}
\item Process sequences of different lengths in a single batch without padding
\item Maintain exact event boundaries without wasted computation
\item Leverage optimized kernels for variable-length attention and SSM operations
\end{itemize}

The nested tensor construction ensures that each sequence in a batch can have a different number of events while maintaining computational efficiency.

\subsection{Model Architecture Details}

\begin{table}[h]
\centering
\caption{Architecture specifications for different model sizes}
\begin{tabular}{lccc}
\toprule
\textbf{Component} & \textbf{Small} & \textbf{Base} & \textbf{Large} \\
\midrule
Hidden dimension & 256 & 512 & 1536 \\
FFN dimension & 1024 & 2048 & 4096 \\
Attention heads & 8 & 16 & 16 \\
Rotary dimension & 32 & 32 & 48 \\
Window size & 3071 & 6399 & -1 \\
SSM state dimension & 32 & 64 & 128 \\
SSM expand factor & 2 & 2 & 2 \\
Total blocks & 8 & 16 & 30 \\
Parameters & 8M & 124M & 1.5B \\
\bottomrule
\end{tabular}
\end{table}

\end{document}